\input{aipcheck}

\documentclass[
    ,final            % use final for the camera ready runs
  ]
  {aipproc}

\layoutstyle{6x9}
\usepackage{bm}

\begin{document}

\title{Symmetry Breaking in Bose-Einstein Condensates}

\classification{03.75.Lm, 67.40.Db, 03.75.Kk, 03.75.Mn}
\keywords      {symmetry breaking, soliton, vortex, dipolar BEC}  

%03.75.Lm 	Tunneling, Josephson effect, Bose-Einstein condensates in periodic potentials, solitons, vortices and topological excitations
%67.40.Db 	Quantum statistical theory; ground state, elementary excitations
%03.75.Kk 	Dynamic properties of condensates; collective and hydrodynamic excitations, superfluid flow
%03.75.Mn 	Multicomponent condensates; spinor condensates

\author{Masahito Ueda}{
  address={Tokyo Institute of Technology, Meguro-ku, Tokyo 152-8551, Japan}
  ,altaddress={Macroscopic Quantum Control Project, ERATO, JST, Bunkyo-ku, Tokyo 113-8656, Japan}
}

\author{Yuki Kawaguchi}{
  address={Tokyo Institute of Technology, Meguro-ku, Tokyo 152-8551, Japan}
}

\author{Hiroki Saito}{
  address={The University of Electro-Communications, Cho-fu, Tokyo 182-8585, Japan}
}

\author{Rina Kanamoto}{
  address={University of Arizona, Tucson, Arizona 85721, USA}
}

\author{Tatsuya Nakajima}{
  address={Tohoku University, Sendai 980-8578, Japan}
}

\begin{abstract}
A gaseous Bose-Einstein condensate (BEC) offers an ideal testing ground for studying symmetry breaking, because a trapped BEC system is in a mesoscopic regime, and situations exist under which symmetry breaking may or may not occur.
Investigating this problem can explain why mean-field theories have been so successful in elucidating gaseous BEC systems and when many-body effects play a significant role.
We substantiate these ideas in four distinct situations: namely, soliton formation in attractive BECs, vortex nucleation in rotating BECs, spontaneous magnetization in spinor BECs, and spin texture formation in dipolar BECs.
\end{abstract}

\maketitle

%%%%%%%%%%%%%%%%%%%%%%%%%%%%%%%%%%%%%%%%%%%%
%% MAINMATTER
%%%%%%%%%%%%%%%%%%%%%%%%%%%%%%%%%%%%%%%%%%%%

\section{Introduction}

Whereas the ground state of a microscopic system possesses the exact symmetry of the Hamiltonian, the ground state of a macroscopic system seldom, if ever, represents the full symmetry of the Hamiltonian. This phenomenon, known as spontaneous symmetry breaking, is the key concept for understanding macroscopic phenomena. 
The purpose of this study is to demonstrate that a gaseous Bose-Einstein condensate (BEC) offers a new paradigm for studying this problem.

Many insights into the nature of superfluidity have been gained through the study of superfluid helium systems. It is therefore instructive to compare the properties of atomic-gas BECs with those of superfluid helium.

One major difference is the kinetics. The collision time for liquid helium is much shorter than the inverse collective mode frequency. This implies that the local thermodynamic equilibrium is ensured and that the physics can be understood by conservation laws and hydrodynamics.
In the case of an atomic-gas BEC, the collision time is of the order of the inverse collective mode frequency. This implies that the local thermodynamic equilibrium cannot always be achieved and that once the system is driven out of equilibrium, the ensuing non-equilibrium relaxation and kinetics are essential for understanding such phenomena as BEC phase transition and vortex nucleation.

Another major difference is symmetry breaking. In the case of bulk liquid helium, the thermodynamic limit is always achieved, and spontaneous symmetry breaking of the relative gauge occurs, resulting in the emergence of the mean field. 
In the case of an atomic-gas BEC, the system is in the mesoscopic regime and the thermodynamic limit is, in general, not achieved. Therefore, whether or not the symmetry breaking occurs depends on the situation of the system, and if it does, its dynamics should be observable due to the aforementioned long collision time.

In the present paper we substantiate these ideas in four distinct cases: namely, soliton formation in quasi one-dimensional attractive BECs, where the translational symmetry is broken; vortex nucleation in rotating BECs, where the axisymmetry is broken; spontaneous magnetization in spinor BECs, where the rotational and chiral symmetries are broken; and the spontaneous formation of spin textures in spinor dipolar BECs, where we can expect the Einstein--de Haas effect and the ground-state mass flow.

%%%%%%%%%%%%%%%%%%%%%%%%%%%%%%%%
\section{Soliton formation in a quasi-1D attractive BEC}

We begin by discussing symmetry breaking in a quasi-1D attractive BEC~\cite{Kanamoto1}. 
Suppose that an attractive BEC is confined in a quasi-1D torus.
On a mean-field level, the properties of the system can be described by the   
Gross-Pitaevskii (GP) equation
\begin{eqnarray}
\left(-\frac{\partial^2}{\partial\theta^2}-
\pi\gamma|\Psi_0|^2\right)\Psi_0=E\Psi_0,
\label{GPS}
\end{eqnarray}
where $\Psi_0(\theta)$ is the ground-state wave function and $\gamma$ is the dimensionless strength of interaction.
Figure 1 shows ground-state wave function $\Psi_0$ as a function of the angular coordinate $\theta$ for various strengths of interaction $\gamma$.

When dimensionless strength of interaction $\gamma$ is smaller than 1, the ground-state density is uniform. However, when it exceeds 1, the translational symmetry is spontaneously broken and a bright soliton is formed as shown in Fig.~1. Thus the mean-field theory predicts a second-order quantum phase transition at $\gamma=1$.
\begin{figure}
  \includegraphics[width=.5\textwidth]{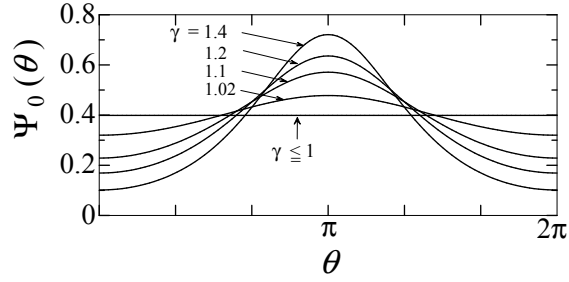}
  \caption{Stationary solutions of Eq.~\eqref{GPS} for several values of $\gamma$. When $\gamma\le 1$, the ground state is a uniform condensate, and when $\gamma$ exceeds 1, the ground state develops into a soliton.}
\end{figure}

Figure 2(a) shows the Bogoliubov spectrum before and after breaking the translational symmetry~\cite{Kanamoto2}. The zero-energy mode emerges in the soliton regime, this being the Goldstone mode associated with the breaking of translational symmetry.
Figure 2(b) shows the corresponding many-body spectrum~\cite{Kanamoto2}.
The many-body spectrum in the uniform BEC regime is similar to that of the Bogoliubov 
spectrum. However, a dramatic change in the landscape of the energy spectrum occurs in the bright soliton regime.
In particular, a quasi-degenerate spectrum appears above the ground state in the bright soliton regime. This quasi-degeneracy is a signature of the breaking of the translational symmetry that generates a bright soliton.
It is remarkable that many-body physics can automatically generate such a     
symmetry-breaking-inducing quasi-degenerate spectrum which is absent  
at the mean-field level.
\begin{figure}
  \includegraphics[width=.9\textwidth]{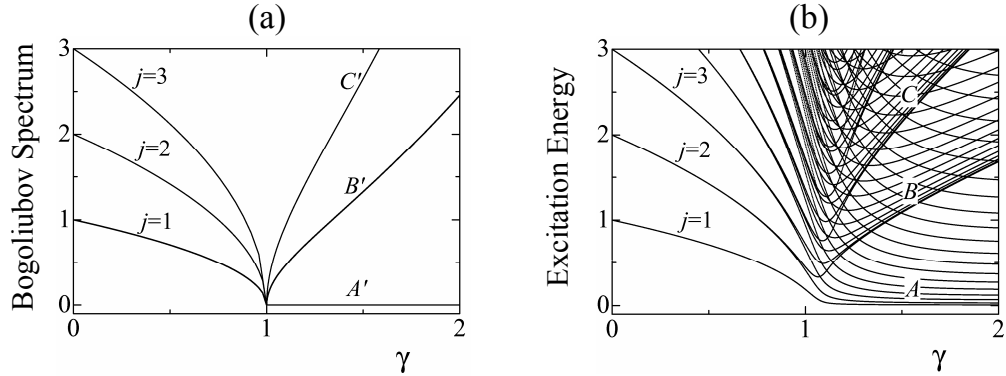}
  \caption{
(a) Bogoliubov spectrum, where branch $A'$ represents the Goldstone mode,
$B'$ the breathing mode of a bright soliton, and $C'$ the second harmonic of $B'$.
(b) Excitation spectrum obtained by exact diagonalization of the Hamiltonian for a system with $200$ particles.
}
\end{figure}

%%%%%%%%%%%%%%%%%%%%%%%%%%%%%%%%%%%%%%%%%%
%\section{Rotating BEC}
\section{Axisymmetry breaking in a rotating BEC}

Let us next discuss symmetry breaking in vortex nucleation.
When the rotational frequency of the container exceeds a certain critical value, the BEC begins to rotate and a vortex enters the system.
The axisymmetry of the system is spontaneously broken in this vortex nucleation process.

Figure 3 shows the many-body excitation spectrum of a rotating BEC as 
a function of the angular momentum $L$ of the system with $N$ particles, in which the energy is measured from the lowest energy state for each angular momentum~\cite{Ueda}.  
The ground state is quasi-degenerate when the rotation frequency is low.
This quasi-degeneracy may be regarded as a precursor for spontaneous symmetry  
breaking due to vortex nucleation.
We note that the energy difference between these quasi-degenerate levels is of the order of $1/N$. This  
quasi-degeneracy is therefore solely of many-body nature and should vanish at the thermodynamic limit.
In fact, we can show that the Goldstone mode associated with the axisymmetry breaking is the lowest-lying envelope of this quasi-degenerate spectrum~\cite{Ueda}.

It is striking that a many-body quasi-degenerate state is spontaneously generated when the continuous symmetry of the system is about to break and then removed when its role is over, i.e., after the symmetry has been broken.
This mechanism for symmetry breaking appears universal, regardless of the
details of the underlying physics, and manifests itself only in the mesoscopic regime.

\begin{figure}
  \includegraphics[width=.45\textwidth]{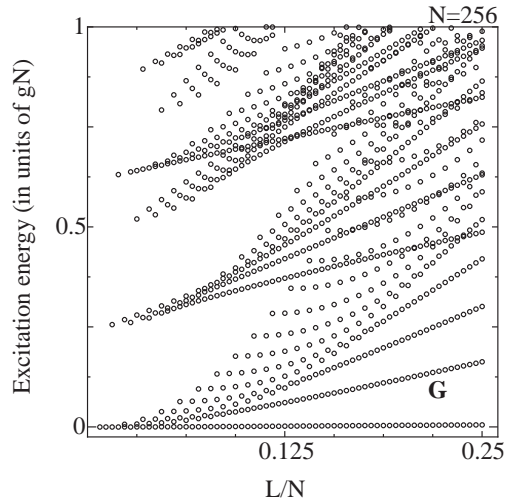}
  \caption{Many-body energy spectrum of a rotating BEC as a function of angular momentum $L$ of a system with $N$ particles.
The energy is measured from the lowest-energy state for each $L$, and excited states that involve center-of-mass motion are not shown. 
The envelope of the lowest-lying excited states that is labeled G represents the Nambu-Goldstone mode associated with axisymmetry breaking. As $L$ increases, 
the envelope develops an energy gap of the order of $N$, indicating mass
acquisition by the Nambu-Goldstone mode~\cite{Ueda}.
}
\end{figure}

%%%%%%%%%%%%%%%%%%%%%%%%%%%%%%%%%%%%%%%%%%%%
\section{Spinor BEC: spin texture and chiral symmetry breaking}

A long-standing question with magnetism is how the spontaneous magnetization of a ferromagnet can occur in an isolated system in which the total spin angular momentum is conserved.
One possible solution is that all spins align in the same direction and that the system is in a quantum-mechanical superposition state over all directions.
We propose here a different scenario, in that the system develops local magnetic domains of various types that depend on the nature of the interaction, conservation laws, and the geometry of the trapping potential.

As an example, let us consider a spin-1 ferromagnetic BEC in a cigar-shaped trap. 
We assume that almost all the atoms are initially prepared in the magnetic sublevel $m=0$.
As time progresses, the $m=0$ population is converted into the $m=\pm1$ modes due to the spin-exchange interaction, and the system develops spin textures. 

Figure 4 shows the time evolution of the mean spin vector on the trap axis~\cite{Saito1}. The length of the spin vector is zero everywhere because of the initial condition.
As time progresses, staggered magnetic domains develop due to dynamical instability.
After more time has elapsed, helical structures are spontaneously formed [Fig.~4(d)] because the kinetic energy is decreased by the formation of a helix. 

\begin{figure}
  \includegraphics[width=.4\textwidth]{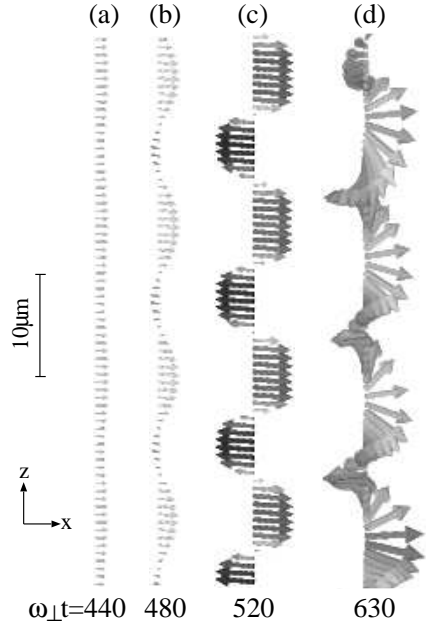}
  \caption{Mean spin vectors along the $z$ axis seen from the $-y$ direction. The trap frequencies are $\omega_\perp = 2\pi \times 245$ Hz and $\omega_z
= 2\pi \times 4$ Hz.
A magnetic field of 54 mG is applied in the $z$ direction.
The initial state is given by $\psi_1 = 0$, $\psi_0 = \sqrt{1 - 10^{-4}} \psi_{\rm g}$,
and $\psi_{-1} = 0.01 \psi_{\rm g}$, where $\psi_{\rm g}$
is the $m = -1$ ground state.
The size of each strip is $600 \times 16$ in units
of $(\hbar / m \omega_\perp)^{1/2} \simeq 0.69~\mu{\rm m}$.
The spin vector is displayed from $z = 0~\mu{\rm m}$ to $z = 54~\mu{\rm m}$ in
(a), (b) and (c), and from $z = 35~\mu{\rm m}$ to $z = 89~\mu{\rm m}$ in (d).
All the vectors are shown in the frame rotating about the $z$ axis at the Larmor frequency.
}
\end{figure}

The ferromagnetic BEC yields yet another surprise: chiral symmetry breaking.
Suppose that all atoms are prepared again in the $m=0$ state in a pancake-shaped trap.
Figure~5 shows the Bogoliubov spectrum as a function of dimensionless strength $g^{\rm 2D}_1 \ (\propto N)$ of the spin-exchange interaction~\cite{Saito2}.
In the region of $g^{\rm 2D}_1<-3.9$, the modes with orbital angular momentum $\ell=\pm1$ have imaginary parts, and they are therefore dynamically unstable and grow exponentially.
When a $^{87}$Rb BEC is prepared in this region,
the $m=0$ atoms are transfered into the $m=\pm1$ states due to the dynamical instability,
and they obtain the orbital angular momentum.
The angular momentum conservation implies that the $m=1$ and $m=-1$ components must have opposite sign of orbital angular momentum.
There are two possibilities: the $m=1$ component can have either orbital angular momentum of $\ell=1$ or $\ell=-1$; 
correspondingly, the $m=-1$ component can have either $-1$ or 1 orbital angular momentum. 
These two possibilities are degenerate, this degeneracy being a statement of the chiral symmetry.
However, since a chirally symmetric state has higher energy than a chiral-symmetry-broken states~\cite{Saito2},
the chiral symmetry is dynamically broken and each spin component begins to rotate spontaneously. 

Figure 5(b) shows the time development of the orbital angular momentum of the $m=-1$ component~\cite{Saito2}.
This remains zero for a certain latency period and then acquires a non-zero value due to the chiral symmetry breaking.
Figures 5(c) and (d) show the spin textures for chirally symmetric and broken symmetry states~\cite{Saito2}.
The chirally symmetric state has a domain wall at the cost of the ferromagnetic energy, while the chiral-symmetry-broken state circumvents this energy cost by developing topological spin textures. This represents the underlying physics of this chiral symmetry breaking.

Our predictions have recently been observed by the Berkeley group~\cite{Stamper-Kurn}.
They carried out experiments subject to the same initial conditions, i.e., all the atoms were initially prepared in the $m=0$ state.
As we predicted, the system remained unmagnetized during a certain latency period, before spontaneously developing magnetization. 
They also observed a polar-core vortex corresponding to our chiral-symmetry-broken state.

\begin{figure}
  \includegraphics[width=\textwidth]{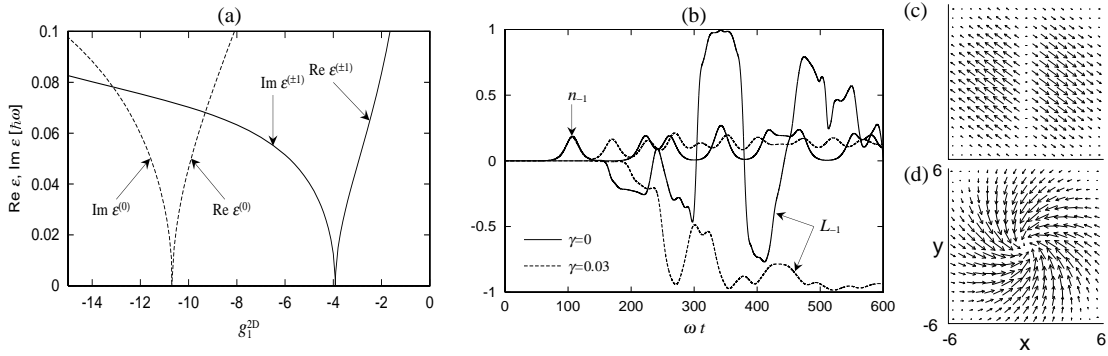}
  \caption{(a) Real and imaginary parts of lowest Bogoliubov energies
$\varepsilon^{(\ell)}$ for $\ell = 0$ and $\pm 1$, where label $\ell$
indicates that the $m = \pm 1$ components of the eigenfunction are
proportional to $e^{\pm i \ell \phi}$ with $(r,\phi)$ being the cylindrical coordinates in two-dimensional system.
The two energy levels of $\varepsilon^{(\pm 1)}$ are degenerate due to the
axisymmetry of the system.
We have taken the parameters of spin-1 $^{87}{\rm Rb}$ atoms, where 
spin-independent interaction strength $g_0^{\rm 2D}$ is related to 
spin-dependent strength $g_1^{\rm 2D}$ by $g_0^{\rm 2D} = -216.1 g_1^{\rm 2D}$.
(b) Time development of the fraction $n_{-1}$ and
orbital angular momentum per particle $L_{-1}$ in the
$m = -1$ component with ($\gamma = 0.03$, dashed) and without ($\gamma = 0$, solid) dissipation.
The interaction strengths are $g_0^{2D}=2200$ and $g_1^{2D}=-10.18$.
The initial state is given by $\psi_0 = \psi_g$, $\psi_{-1} = 10^{-4} r
(e^{i \phi} + 1.0001 e^{-i \phi}) \psi_g$, and $\psi_1 = 0$, where
$\psi_g$ is the $m=0$ mean-field ground state.
(c) and (d) show snapshots of the spin textures.
The size of the frame is $12 \times 12$ in units of $a_{\rm ho}$.
}
\end{figure}

%%%%%%%%%%%%%%%%%%%%%%%%%%
\section{Dipolar BEC: Einstein--de Haas Effect and ground-state mass flow}

Let us finally discuss a dipolar BEC by focusing on the spin degree of freedom.
The magnetic dipole-dipole interaction is a tensor force which causes spin-orbit coupling,
and only the total, spin plus orbital, angular momentum is conserved.
Therefore, the dipolar interaction transfers angular momentum between the spin and orbit,
i.e., the Einstein--de Haas effect occurs.

%%%%%%%%%%%%%%%%%%%%%%%%%%%%%%%%%%%%%%%%%%%
The Stuttgart group created a spin-polarized $^{52}$Cr BEC under a relatively strong magnetic field~\cite{Pfau}.
What would happen to this BEC if we reduced the magnetic field to zero?
We could expect spin relaxation to occur, and because of the conservation of total angular momentum, BEC would start to rotate spontaneously.
This is the Einstein--de Haas effect~\cite{Kawaguchi1,Santos}.
To check this idea, we performed numerical simulations by solving the seven-component non-local GP equation in three dimensions.
We find that the BEC begins to rotate spontaneously to compensate for a
decrease in the spin angular momentum, as shown in Fig.~6(a).
We also find that atomic spins undergo Larmor precession around the local dipole field, and develop topological spin textures as shown in Fig.~6(b) and (c). 
In this case, the dipole field is directed inward in the upper hemisphere and outward in the lower hemisphere, and we therefore have spin textures with opposite directions. 

\begin{figure}
  \includegraphics[width=.8\textwidth]{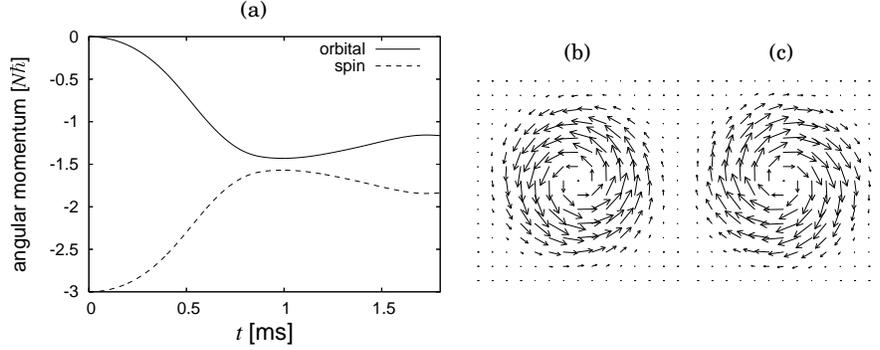}
\caption{(a) Time development of the spin and orbital angular momenta in a spherical trap in the absence of a magnetic field.
The initial state is spin polarized at $m=-3$.
(b) and (c) show snapshots of the spin textures on (b) the $z=2~\mu$m plane and (c) the $z=-2~\mu$m plane at $t=0.4$ ms,
where the Thomas-Fermi radius is $3.5~\mu$m.
The length of the arrow represents the magnitude of the spin vector projected on the $x$--$y$ plane.
}
\end{figure}

%%%%%%%%%%%%%%%%%%%%%%%%%%%%%%%%%%%%%%%%%%%%
The dipole-dipole interaction is also expected to yield ground-state spin texture in a ferromagnetic BEC,
as in the case of the domain structure in a solid-state ferromagnet.
The Einstein--de Haas effect is also known to occur in a ferromagnet.
Then, what is truly new in spinor dipolar BECs?

The unique feature of the spinor dipolar BEC which is absent from a solid-state ferromagnet is the spin-gauge symmetry which relates the spin texture to a mass current.
The fundamental query is whether or not a spinor dipolar BEC can exhibit a spontaneous mass current in the ground state.
The answer is summarized in the phase diagram shown in Fig. 7(a)~\cite{Kawaguchi2}, in which $R_{\rm TF}$ is the Thomas-Fermi radius, $\xi_{\rm dd}$ is the dipole healing length, and $\xi_{\rm sp}$ is the spin healing length.

We have found three phases.
When the system size is small, the flower phase is stabilized in which
spins are almost polarized and slightly flare out depending on $R_{\rm TF}/\xi_{\rm dd}$.
When the dipolar interaction becomes stronger, or equivalently the system size becomes larger,
the spin vectors tilt into the either $\phi$ or $-\phi$ direction, where $\phi$ is the azimuthal angle.
This spin texture has chirality and we call this phase the chiral spin-vortex phase.
In this phase, BEC has a substantial net mass current which monotonically increases as a function of $R_{\rm TF}/\xi_{\rm dd}$,
leading to increase of the kinetic energy.
Therefore, when $R_{\rm TF}/\xi_{\rm dd}$ becomes larger, the system goes to the next phase: the polar-core vortex phase,
the spin texture in this phase forming a vortex with no net circulation.

All three phases can be realized in the spin-1 $^{87}$Rb BEC by varying the trap frequency.
Figure 7(b) shows the orbital angular momentum of the ground-state spin-1 $^{87}$Rb BEC as a function of the trap frequency and the number of atoms.
We find particularly in the chiral spin-vortex state, that the orbital angular momentum is increased by up to 40 \% of the full value of the singly quantized circulation.

\begin{figure}
 \includegraphics[width=.95\textwidth]{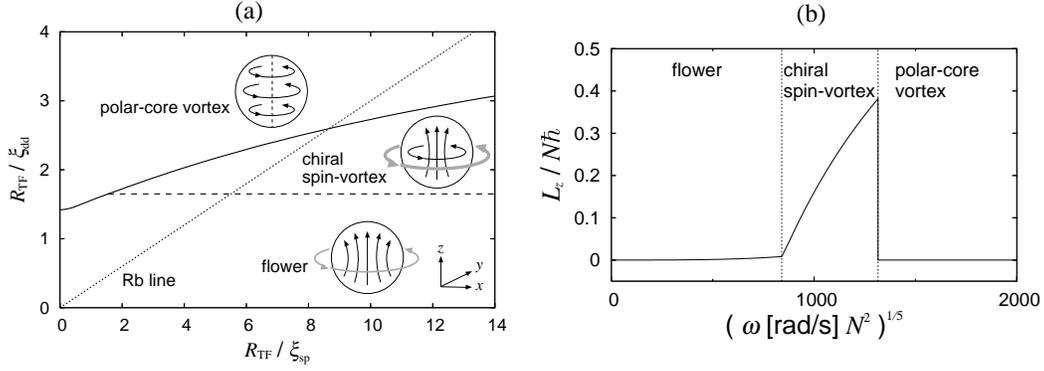}
 \caption{(a) Phase diagram of a spin-1 ferromagnetic dipolar BEC.
The solid curve shows the boundary between the states with total angular momentum $J=0$ and $J=1$,
and the broken line divides the phases with and without spin chirality.
The schematic diagram shown in each region represents the spin configuration (black arrows) and mass flow (gray arrows).
The spin-1 $^{87}$Rb BEC traces the dotted line where $\xi_{\rm sp}/\xi_{\rm dd}=0.30$.
(b) Orbital angular momentum as a function of trap frequency $\omega$ and number of atoms $N$ for a spin-1 $^{87}$Rb BEC along the dotted line in Fig.~7(a).
In the case of a BEC with a million atoms, the chiral spin-vortex phase exists in the region of $2\pi \times 70~{\rm Hz}\le \omega \le 2\pi\times 630~{\rm Hz}$.
}
\end{figure}

%%%%%%%%%%%%%%%%%%%%%%%%%%%%%%%%%%%%%%%%%%%%%
\section{Summary}

A gaseous BEC is a mesoscopic system.
Studying this system enables us to learn in detail when and how symmetry breaking occurs, and to identify the dynamics of symmetry breaking because of the long collision time.
Various types of symmetry breaking are experimentally accessible.
In the case of soliton formation, the translational symmetry is spontaneously   
broken; 
in the case of vortex nucleation, the axisymmetry is broken;
in the case of a spinor BEC, the rotational and chiral symmetries are broken and various spin textures are spontaneously generated due to competition between the conservation law, interaction, and geometry of the system; and 
in the case of a dipolar BEC, we expect the Einstein--de Haas effect and  
substantial ground-state circulation in the chiral spin-vortex phase.

\end{document}